\documentclass[twocolumn]{aastex631}
\usepackage{amsmath}

\newcommand{\editone}{}

\begin{document}

\title{Thermal Evolution and Mass Loss on Short-Period, Low-Mass Planets During FU Orionis Outbursts}

\author[0000-0002-7733-4522]{Juliette Becker}
\affiliation{Department of Astronomy, University of Wisconsin-Madison, 475 N. Charter Street, Madison, WI 53706, US}
\author[0000-0002-7094-7908]{Konstantin Batygin}
\affiliation{Division of Geological and Planetary Sciences, California Institute of Technology, Pasadena, CA 91125, USA}

\begin{abstract}

Ultra-short-period (USP) planets represent a unique class of exoplanets characterized by their tight orbits and relatively low masses, with some also exhibiting unusually high iron fractions. Previous work (Becker et al, 2021) proposed a dynamical pathway wherein planets can migrate inward due to drag from sub-Keplerian gas during episodic FU Orionis (FU Ori) outbursts, an abrupt accretion phenomenon exhibited by young stellar objects, thereby potentially populating USP orbits. However, the implications of this migration process on the structural and compositional evolution of these planets remain unexplored. In this work, we model the response of a planet's surface material to the high disk temperatures characteristic of an FU Ori event and compute the fraction of an Earth-like planet's mass that will be lost due to vaporization and subsequent turbulent diffusion of gaseous molecules during the FU Ori event. We find that low-mass planets may lose a substantial fraction of their mantle mass during FU Ori events, potentially contributing to the observed prevalence of low-mass, iron-rich USP planets.

\end{abstract}



\section{Introduction}
\label{sec:intro}

FU Orionis (FU Ori) outbursts are substantial, rapid increases in the brightness of a young stellar object due to surges in the accretion rate - the rate at which from material from the surrounding protoplanetary disk feeds onto the central star \citep{Herbig1966, Herbig1977}. 
The accretion rates during these outbursts often exceed the typical rates seen in T Tauri stars by several orders of magnitude \citep{Hartmann1996}, which results in considerable structural alterations to the inner disk \citep{Bell1994,Popham1996, Tapia2017,Shu2007, Rodriguez2022}. While the general origin of these accretion outbursts is not yet known, in some cases, it is thought that these events are due to the presence of massive planets driving gaps and subsequently bursts in accretion close to the star \citep{Nayakshin2012, Nayakshin2023, Nayakshin2024}. 

In our previous work \citep{Becker2021}, we showed that sub-Jovian planets ($<4M_{\oplus}$) residing close to the inner edge of the protoplanetary disk can migrate inwards during an FU Ori outburst and become trapped within the disk's magnetospheric cavity. 
In concert with such episodic accretion events, the disk edge decreases in radius to nearly the stellar radius.
If gas in the protoplanetary disk orbits at a sub-Keplerian velocity due to strong magnetic fields, it can provide a strong headwind torque to low-mass planets, allowing them to overcome the locally outward Type I migration torque.  
This mechanism provides another avenue \citep[in addition to other pathways involving planet-planet interactions proposed in][]{Lee2017, Petrovich2019, Serrano2022} to place planets in ultra-short-period orbits.

\editone{Ultra-short-period (USP) planets are defined as those with orbital periods of roughly a day or less \citep{Winn2018, Goyal2025}. They} are relatively rare, orbiting only $\sim$0.5\% of stars \citep{SanchisOjeda2014}, placing them in a distinct parameter space compared to the rest of the exoplanet sample. 
The radii of USP planets tend to be less than 2 $R_{\oplus}$ \citep{Winn2018}, and 
a subset of USP planets exhibit remarkably high bulk densities implying sizeable iron fractions \citep[ex: KOI 1843.03, GJ 367 b, TOI-1075 b;][]{Price2020, Lam2021, Essack2023}.
Given their close proximity to their host stars, USP planets and their short-period neighbors reside within regions of the protoplanetary disk that can reach extraordinarily high temperatures \citep{Zhu2020}, especially during periods of FU Orionis (FU Ori) outbursts. 
In this work, we use a simple physical model to consider how the high disk temperatures during FU Ori events could drive mass loss in rocky planets near the disk edge. 
In Section \ref{sec:disk}, we construct a simple model that describes the response of a planet's surface material during an FU Ori outburst event. Specifically, we consider how typical FU Ori disk temperatures may heat the surface of a planet to a level where the rock vaporizes. We then, in Section \ref{sec:turb}, discuss the conditions required for this material to subsequently escape the planet's gravitational influence (which is necessary to avoid re-condensation after the FU Ori event is over). Finally, in Section \ref{sec:discuss} we will conclude and discuss the implications of the results.

\section{The Physical Model of Planet Vaporization}
\label{sec:disk}
In this section, we will build a simple model to approximate the vaporization of a planet's surface in a hot disk environment. 
We begin by defining the temperature profile of the circumstellar disk during an FU Ori outburst event following \citet{Rodriguez2022}:
\begin{equation}
T(r) = \begin{cases}
  T_{max} \left( \dfrac{r}{1.36 R_{*}}\right)^{\gamma} &\hspace{-15mm} \rm{for\ } R_{*} < r < 1.36 R_{*} \\
  \left(\dfrac{3 G M_{*} \dot{M}}{8 \pi \sigma r^3} \left(1 - \sqrt{R_{*}/r}\ \right)\right)^{1/4}  & \rm{for\ } r > 1.36 R_{*} 
\end{cases}
\label{eq:tempFUori}
\end{equation}
where $R_{*}$ is the stellar radius, $r$ is the radius at which the temperature profile is being evaluated, $T_{max} = 5700$ K is the disk temperature at a distance $r_{0} = 1.361\ R_{*}$ AU. 
It is important to note that the temperature profiles of different disks will vary (in $T_{0}, r_{0}$, and $\gamma$), resulting in some slight dispersion of the temperatures experienced by a planet at a given orbital radius. In this work, we use $\gamma = 0$ as in \citet{Rodriguez2022} for definiteness (see also \citealt{Labdon2021}). 

\begin{figure}
 \includegraphics[width=3.3in]{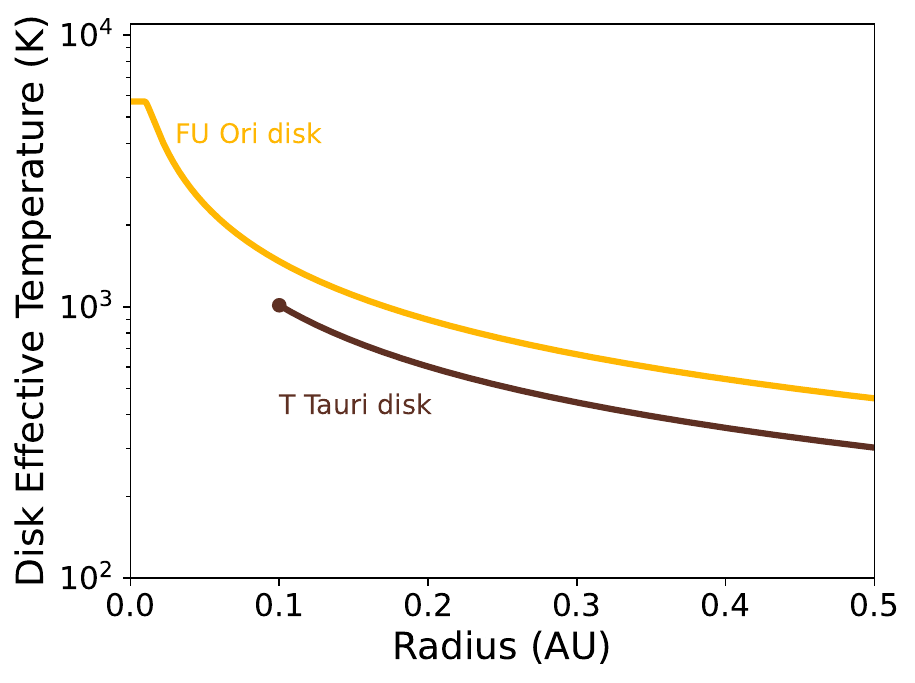}
 \caption{The disk effective temperature profiles for the typical T Tauri disk and the FU Ori disk. During FU Ori events, not only does the disk edge approach the surface of the star, but the disk temperature is significantly increased at short radii.  }
    \label{fig:fig1}
\end{figure}

As a comparison, a typical flat, thin T Tauri disk has a temperature profile of the form \citep{Adams1988}:
\begin{equation}
T(r) = T_{0} (r_{0} / r)^{q},
\label{eq:tempdistribution}
\end{equation}
where $r$ is the radius at which the temperature profile is being evaluated, $T_{0} = 180$ K is the disk temperature at a distance $r_{0} = 1$ AU, and $q = -0.75$ the power law index of the temperature profile. The disk is truncated at 0.1 AU \citep{Batygin2023}, resulting the maximum disk temperature being roughly 1000 K at the inner disk edge. 
Figure \ref{fig:fig1} shows a comparison between the disk temperature profiles during an FU Ori outburst event and during the quiescent T Tauri phase. 

Provided that planet formation and migration can ensue on a timescale short enough that FU Ori outbursts still occur after planets have reached the inner disk edge \citep[ex:][]{Batygin2023}, the dramatic changes in disk conditions triggered by an FU Ori outburst event profoundly affect the thermal conditions of these planets. 
Disk midplane temperatures, where embedded planets may reside, may increase by an order of magnitude of more during FU Ori events. 
Specifically, these cataclysmic events can cause the disk midplane temperature surrounding such planets to surge to as high as $3  - 6\times 10^{4}$ K \citep{Zhu2018, Zhu2020}. 
The exact temperature profile of the disk is sensitive to initial conditions including magnetic field strength \citep{Zhu2018}, shock temperature and dynamics \citep{Carvalho2024}, and ionization state \citep{Nayakshin2024b}, among other effects.

This heightened thermal energy subsequently permeates the embedded planets, raising their atmospheric temperatures to a temperature commensurate with the local disk temperature. 
These extreme temperatures exceed the 1 bar sublimation point (3690 K) of forsterite (Mg$_{2}$SiO$_{4}$), which is a predominant constituent of Earth's mantle. As a result, rocky material on the surface will be heated, melted, heated again, and then vaporized by the excess heat provided by molecular collisions. 

At the surface-atmosphere boundary, the temperature gradient is set by molecular diffusion. 
Once sublimated, molecules will diffuse from the surface (which will have a high concentration of forsterite vapor including Mg, O, O$_{2}$, SiO, MgO, Si$_{2}$; \citealt{Fegley2012}) towards the lower concentration. 
At the planetary surface, the temperature is equal to the boiling point temperature of forsterite, $T_{\rm{FO}} = 3960$ K, while above the interface of the vaporization the temperature is equal to the protoplanetary disk temperature T. 
Assuming molecular diffusion in the boundary layer, the temperature gradient over this interface can be written as:
\begin{equation}
\dfrac{dT}{dz} \approx \dfrac{\Delta T}{\sqrt{\nu_{M} \tau}},
\end{equation}
when $\nu_{M} = c_{s} \mu m_{H} / \rho a$ is the molecular viscosity, $a$ the molecule radius, $\mu$ the mean molecular mass in amu, $m_H$ the mass of the hydrogen atom, $\Delta T = T - T_{FO}$ the temperature difference between the atmosphere temperature and the boiling point of the rocky material, and $\tau$ the elapsed time of the FU Ori event. 
The density $\rho$ of gas at the planetary surface, which is needed to compute the molecular viscosity, can be estimated by computing the planetary atmosphere density profile. 

In a protoplanetary disk, a planet embedded in the gaseous disk will have a hydrostatic envelope. The extent of this envelope can be quantified by the Bondi radius ($r_{B}$) \editone{or the Hill radius ($R_H$)}, which \editone{are measures of the gravitational influence of the planet on the surrounding gas. For hot disk material, $R_b< R_H$ and the envelope size will be set by the Bondi radius}. $R_b$ is expressed as:
\begin{equation}
R_{B} = \frac{G m_{p}}{c_{s}^{2}},
\label{eq:bondi}
\end{equation}
where $G$ denotes the gravitational constant, $m_{p}$ represents the mass of the planet, and $c_{s}$ stands for the sound speed of the surrounding gas.
The speed of sound $c_s$ is set by the temperature ($T$) and molecular mass ($\mu$) of the local gas:
\begin{equation}
c_{s} = \sqrt{\frac{k_{B}T}{\mu_{d} m_{H}}}.
    \label{eq:speedofsound}
\end{equation}
The mean molecular weight of the protoplanetary disk is roughly $\mu_{d} = 2.4$ amu \citep{Kimura2016}. 
\editone{In comparison, $R_H$ is set only by gravitational considerations and will set the size of the planetary atmosphere for cooler discs and for planets at smaller semi-major axis values. $R_{H}$ will be written as:}
\begin{equation}
R_{H} = a \Bigg(\frac{M_{p}}{3 M_{*}} \Bigg)^{1/3}
\label{eq:bondi}
\end{equation}
\editone{where $a$ denotes the planetary semi-major axis.}

The disk surface density profile is set by:
\begin{equation}
\Sigma(r) = \zeta \Sigma_{0} (r_{0} / r)^{p},
\label{eq:surfdens}
\end{equation}
where $\Sigma_{0}$ is the surface density at distance $r_{0}$, and $p$ the power law index (which is typically taken to be 3/2). We also define $\zeta =0.1$, a dimensionless scaling factor that determines how the bulk density near the disk edge changes in response to the high-accretion rates typical of FU Ori outbursts. 
For a typical disk, $\Sigma_{0}$ = 20000 kg/m$^2$ at $r_{0}$ = 1 AU. We also choose $\zeta = 0.1$, and $h/r = 0.03$.
These values determine an important boundary condition: at the Bondi radius, the planetary atmosphere density will become equal to the density of the protoplanetary disk ($\rho(R_b) = \rho_d$). 
With these parameters defined, we can write the expression for $\rho(R_b)$:
\begin{equation}
\rho(R_b) = \rho_d =  \Sigma(r) / (h\sqrt{2 \pi})
\end{equation}
which can be evaluated at the typical location of an USP planet, $r = 0.02$ AU, to give a estimate for the density that varies with the temperature of the disk: $\rho \approx 0.003$ kg/m$^3$, commensurate with the 0.0001 kg/m$^3$ midplane density used in the numerical simulations of \citet{Zhu2020}.

\begin{figure}
 \includegraphics[width=3.3in]{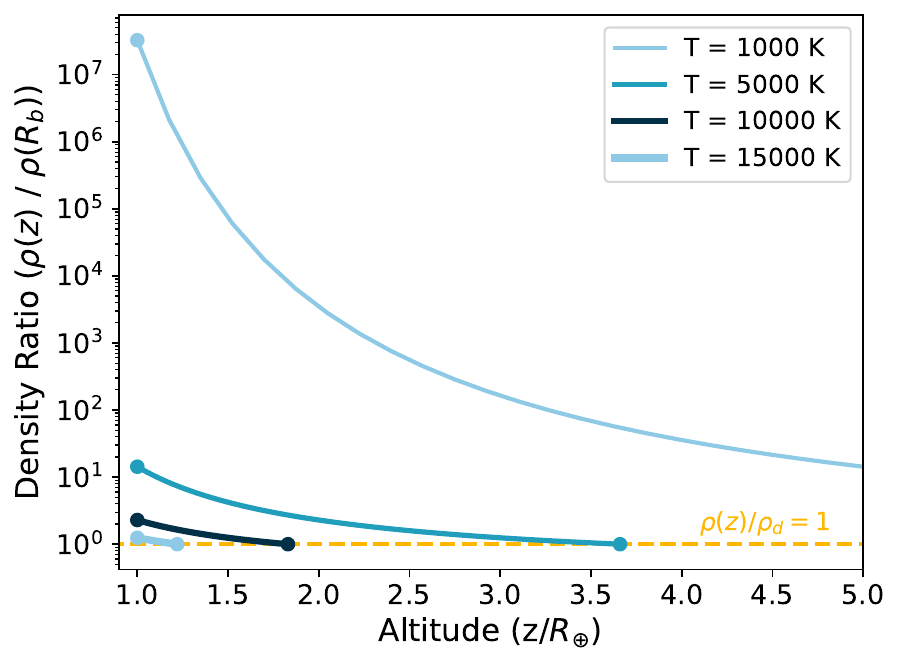}
 \caption{The density of the planetary atmosphere as a function of the altitude in units of the planet radius \editone{for four different disk temperatures}. At the Bondi radius, material is no longer gravitationally bound to the planet and the density becomes equal to the disk local bulk density. \editone{For disk temperatures above 10000 K, the Bondi radius approaches the planetary radius.  } }
    \label{fig:fig3}
\end{figure}

{In constructing our model, we assume a hydrostatic envelope for the planet embedded in the gaseous disk. This assumption is supported by recent simulations \citep{Bailey2023}, which suggest that for planets embedded in protoplanetary disks, a significant portion of the proto-envelope maintains hydrostatic equilibrium. Consequently, to}
determine the density profile of the planetary atmosphere, we solve the equation of hydrostatic equilibrium, $dP/dr = -\rho g$, assuming an isothermal equation of state

\begin{equation}
\frac{k_b T}{\mu m_H} \frac{d \rho(z)}{dz} = -\rho \frac{G M}{z^2},
\label{eq:hse}
\end{equation}
which yields \editone{the solution}
\begin{equation}
\rho(z) =  \rho_{d} \exp \Bigg(\frac{G M \mu m_H}{k_b T}\bigg(\frac{1}{z} - \frac{1}{R_b}\bigg)\Bigg).
\label{eq:rho}
\end{equation}
For an embedded Earth-like planet, the Bondi radius \editone{evaluates to $R_b \approx 3.6 R_{\oplus}$,} provided our chosen values for the FU Ori outburst. 
In Figure \ref{fig:fig3}, we show the density profile for the embedded planet before any rocky surface material has been lost, scaled by the density at the Bondi radius $\rho(R_b) = \rho_d$. 

To determine the heat flux at the planet's surface, we assume radiative heat transfer equation at the boundary layer, and substitute in Equation \ref{eq:tempdistribution}, which yields
\begin{equation}
F \approx \dfrac{\sigma T_{\rm{FO}}^{3}}{\kappa \rho} \frac{dT}{dz} \approx \dfrac{\sigma T_{\rm{FO}}^{3}}{\kappa \rho} \dfrac{\Delta T}{\sqrt{\nu_{M} \tau}},
\label{eq:diffuse}
\end{equation}
when $\sigma$ is the Stefan-Boltzmann constant. 
For a boundary condition temperature of $T = 3690$ K at the surface of the planet where material is being vaporized, the opacity will be approximately $\kappa = 0.01$ m$^{2}$/kg\footnote{It is important to note that this parameter is very sensitive to chemical and temperature conditions and will vary between planets.} \citep{Pollack1985, Vazan2013}.  
The heat flux into the planet will determine the energy available to heat, melt, and evaporate rock from the surface of the planet, which we assume to be made primarily of forsterite. 
The enthalpy of vaporization of forsterite can be computed, as in \citet{Fegley2012}, as the sum of the enthalpies required to heat forsterite to its melting point (330 kJ/mol), to melt solid forsterite (114 kJ/mol), to heat molten forsterite to its vaporization point (313 kJ/mol), and finally to vaporize forsterite into dissociated gases (430 kJ/mol). The total enthalpy of reaction is roughly $H_{vap}\approx$ 1200 kJ. The molecular flux removed from the planet by heat flux $F$ can be computed by:
\begin{equation}
\frac{dN_{FO}}{dt} \approx F * (4 \pi R_{p}^2) / H_{vap}
\end{equation}
when $N_{FO}$ is the number of moles of forsterite vaporized from the planet surface. This can be converted into an amount of mass loss over a single FU Ori event lasting a timescale $\tau$ years:
\begin{equation}
M_{lost} \approx \tau \frac{dN_{FO}}{dt} \mu_{FO} m_{H}\approx \mu_{FO} m_H \tau F * \frac{4 \pi R_{p}^2}{H_{vap}}
\label{eq:mlost}
\end{equation}
The value computed by Equation \ref{eq:mlost} is the total amount of mass lost over one $\tau \sim 100$ yr FU Ori outburst event. 

Due to the uncertainty in the disk temperature, we present the results of our mass-loss calculation as a function of the temperature experienced by the planet. For context, the effective temperatures in these systems (as shown in Figure \ref{fig:fig1}) are approximately 3000 - 6000 K during FU Ori outbursts in the region of the disk that we examine, but the exact disk effective temperature varies depending on the exact radius at which the planet resides, the accretion rate of the disk material onto the central star, and other parameters. Furthermore, the planet will likely experience significantly hotter disk temperatures near the disk midplane compared to the observed effective temperature, 
with the temperatures of the disk midplane being potentially up to $5\times 10^{4}$ K or so \citep{Zhu2020}.
As a result, there is significant uncertainty in the exact temperature experienced by a typical planet in the disk locations that we are considering. 

\begin{figure}
 \includegraphics[width=3.3in]{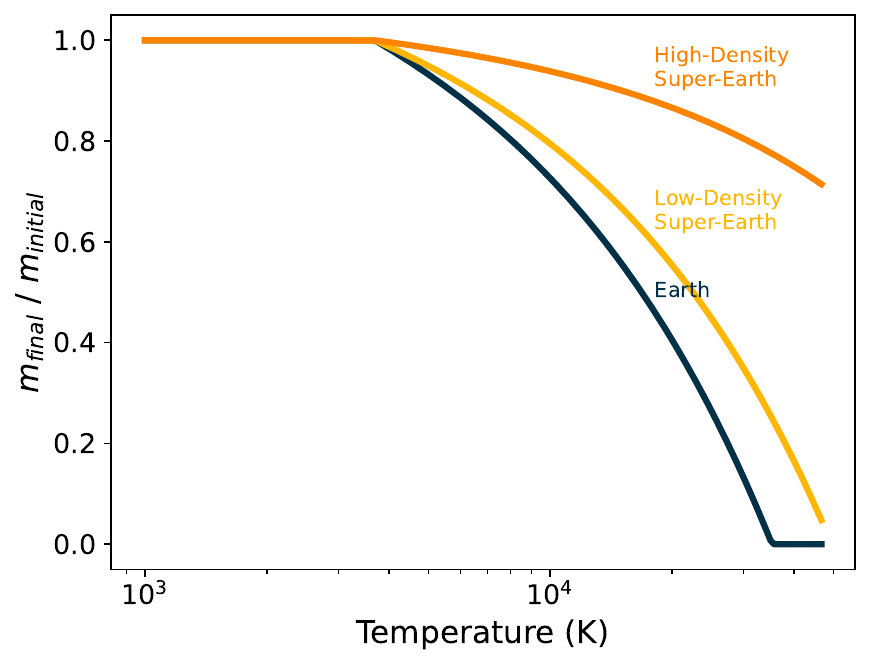}
 \caption{The ratio of the final mass of the planet after one FU Ori outburst event to the initial planet mass, plotted as it depends on the local disk temperature. Three representative planets are plotted: an Earth-like planet ($1 M_{\oplus}$, $1 R_{\oplus}$), a low-density super-Earth ($3 M_{\oplus}$, $1 R_{\oplus}$), and a high-density super-Earth ($10 M_{\oplus}$, $1 R_{\oplus}$). The planets were chosen to span the range of densities seen at the boundaries of the super-Earth ($R_p < 1.5 R_{oplus}$) population. 
 For temperatures above the boiling point of forsterite, the non-gaseous planet mass will decrease as material is vaporized and released into the planetary atmosphere in a gaseous form.  In our toy model, Earth-like planets are fully vaporized in 100 years at the higher range of possible temperatures, while high-density super-Earths will survive at all temperatures. }
    \label{fig:fig2}
\end{figure}

In Figure \ref{fig:fig2}, we plot how the ratio of the initial planet mass to final planet mass (as computed by Equation \ref{eq:mlost}) depends on the local physical disk 
temperature. We show three planet structures: an Earth-like planet ($1 M_{\oplus}$, $1 R_{\oplus}$), a low-density super-Earth ($3 M_{\oplus}$, $1 R_{\oplus}$), and a high-density super-Earth ($10 M_{\oplus}$, $1 R_{\oplus}$). 
For our tested combination of parameters (disk parameters, planet parameters, and atmospheric opacity), we find that an Earth-like planet would be fully lost within one typical FU Ori outburst at the larger expected disk midplane temperatures ($4-5\times 10^{4}$ K). At such high temperatures, only the most massive super-Earths would be expected to survive. 
While the exact planet masses that survive in Figure \ref{fig:fig2} would vary significantly beyond the range shown in the figure with each planet’s and disk’s specific parameters, this calculation highlights that lower-mass planets are at particular risk of being lost.

\section{Turbulent Recycling of the Atmosphere}
\label{sec:turb}
Once the planet's surface material has sublimated, the resulting gaseous molecules will persist in the planet's atmosphere \editone{if the planet retains a distinct atmosphere.
As shown in Figure \ref{fig:fig3}, for disk temperatures above $10^4$ K, the Bondi radius approaches the planet's physical radius, while for cooler disk temperatures the planet may have an atmosphere distinct from the disk material. In FU Ori disks, disk conditions will generally lead to planets being emersed in disk material with temperatures above $10^4$ K. }

\subsection{\editone{High-Temperature Disk Regions}}
The \citet{Shakura1973} parametrization introduces $\alpha$, a dimensionless parameter related to the Mach number such that $ v_{turb} = \sqrt{\alpha} c_s$. Similarly, the length scale of turbulence can be written as $l_{turb} =  \sqrt{\alpha} h$ where $h$ is the scale height.
\editone{In high-temperature disk regions in FU Ori disks, $h/r \sim 0.3 - 0.5$ \citep{Bell1994}.} In disks, typical values of $\alpha$ during FU Ori outbursts are estimated to be $\alpha \approx$ 0.02 - 0.2 \citep{Zhu2007}, compared to $\alpha \approx 10^{-4} - 0.04$ \citep{Rafikov2017} for the general disk sample. 
\editone{This means that a disk with $h/r = 0.3$, $\alpha = 0.1$, and at a distance of 0.02 AU would have a turbulent length scale $l_{turb} \approx 0.002$ AU $ \approx 45 R_{\oplus}$. An embedded planet would thus be physically much smaller in size than the turbulent length scale of the disk.}

In the context of turbulent transport, the Reynolds decomposition approach allows us to express turbulent diffusion as a product of a turbulent length-scale and a turbulent velocity scale, a method conventional in modeling sub-sonic turbulence.
In disks, it is conventional to parameterize turbulent viscosity as a product of the local typical turbulent velocity and the typical turbulent length scale
\begin{equation}
    \mathcal{D}_{turb} = v_{turb} l_{turb}.
\end{equation}
We can re-write the diffusion coefficient as 
\begin{equation}
    \mathcal{D}_{turb} = \alpha c_s h = \alpha \frac{c_s^3}{g}.
    \label{eq:dturb}
\end{equation}
when $g$ is the vertical component of the gravitational acceleration near the planet.
This approximation holds for objects over masses ranging from planetary to stellar \citep{Burrows2001, Marley2015, Sonoi2019}.

We also can establish a threshold corresponding to the point at which the planetary material will diffuse outwards to the planet's Bondi radius within the timescale of the FU Ori event ($\tau$). This threshold will occur when the turbulent diffusion coefficient is equal to the value required to transport material over a length scale of the Bondi radius over a timescale of the FU Ori outburst time $\tau$: 
\begin{equation}
\mathcal{D}_{turb,crit} = \frac{R_b^2}{\tau}
\label{eq:dturb2}
\end{equation}
Upon reaching the Bondi radius, the gaseous planetary material is permanently removed from the planet's gravitational influence and subsequently integrates into the broader disk environment.

We can equate these two previous expressions to place a limit on the level of viscosity required to transport material from the planetary surface to the Bondi radius within the timescale $\tau$:
\begin{equation}
\alpha \frac{c_s^3}{g} > \frac{R_b^2}{\tau}
\label{eq:setup}
\end{equation}
which yields the constraint
\begin{equation}
\alpha > (g R_b^2) / (\tau c_s^3).
\label{eq:alpha_limit}
\end{equation}
However, \editone{as shown in Figure \ref{fig:fig3}, in high temperature disk regions the Bondi radius of an embedded planet will approach the planet's physical radius, such that $R_b = R_p$. In this case, and since $g = G M_p / R_p^2$ at the planet's surface, we can rewrite Equation \ref{eq:alpha_limit} as: }
\begin{equation}
\alpha > (G M_{p}) / (\tau c_s^3).
\end{equation}
 
\editone{For a $4 M_{\oplus}$ planet embedded in a disk region with a temperature $T = 15000$ K for an outburst length of 100 years, this yields a viscosity constraint of $\alpha > 1 \times10^{-6}$ for all planet mass to be lost. Even if the outburst event were only 1 or 10 years, this limit would only reduce by a factor of 10-100, and would still reside well below the predicted disk viscosity values of FU Ori disks of $\alpha \approx$ 0.02 - 0.2 \citep{Zhu2007}.}

When $\alpha$ is above this threshold value, \editone{which will be expected in nearly all FU Ori disks}, the strength of turbulence will be sufficient to cause the material vaporized from the planet's surface be mixed with the disk material \editone{and lost permanently from the planet. For disks in the temperature range $T>3 \times 10^4$ K, Earth-sized planets will likely be fully vaporized}.

\subsection{\editone{Low-Temperature Disk Regions}}
\editone{In disk regions with low temperatures ($5 \times 10^{3} < T<10^4$ K or so), the assumption $R_p = R_b$ which was used to simplify Equation \ref{eq:alpha_limit} no longer holds. In addition, at these temperatures and in the range of $a = 0.01 - 0.02$ AU, the Hill radius $R_{H} < R_B$ and will set the distance of escape for atmosphere material. }

\editone{In this case, we can rewrite Equation \ref{eq:setup} to use the Hill radius as the escape radius:}
\begin{equation}
\alpha \frac{c_s^3}{g} > \frac{R_H^2}{\tau}
\label{eq:setup2}
\end{equation}
which, evaluating $g$ at the planetary radius, simplifies to:
\begin{equation}
\alpha > \frac{G\,a^2\,M_p^{5/3}}{c_s^3 \,\tau \,R_p^2\,\bigl(3\,M_*\bigr)^{2/3}}
\label{eq:final2}
\end{equation}

\editone{For a 4 $M_{\oplus}$ planet at 0.02 AU in a disk region with temperature 3000 K, the viscosity must be $\alpha \gtrsim 0.001$ in order for the material to be transported to the Hill radius and lost.} 
Even in the case of a planet embedded in a cooler disk where a distinct planetary atmosphere is still present, it is reasonable to anticipate that turbulent diffusion of material in the planetary atmosphere will proceed at a rate driven by the local disk region's $\alpha$. 

Although observational measurements of $\alpha$ for planetary atmospheres are limited, recent JWST observations of WASP-107b \citep{Sing2024}, a planet with a temperature of about 500~K, revealed an atmospheric diffusion coefficient of $D = 10^{11.6 \pm 0.1}\,\mathrm{cm}^2\,\mathrm{s}^{-1}$.
This value corresponds to an estimated viscosity of $\alpha \approx 10^{-2}$. \editone{Similar measurements have not yet been made for significantly hotter planets than this, and we assume here that planets embedded in FU Ori disks will inherit the $\alpha$ of the disks. }


\section{Discussion}
\label{sec:discuss}

In this paper, we provide a mechanism by which planets residing near the inner edge of the protoplanetary disk can lose substantial fractions of their planetary mass during FU Ori outburst events. In particular, this mass loss will affect planets with the shortest orbital periods (USP planets). 
USP planets tend to be smaller ($<2R_{\oplus}$; \citealt{Winn2018}) than the exoplanet sample as a whole, and while most of them do not have direct mass constraints, there are multiple systems with such measurements \citep[for example,][]{Price2020, Lam2021, Essack2023, Livingston2024} that show remarkably high bulk densities \editone{potentially} indicating high iron fractions. 

The mechanism we propose in this work, invoking the loss of a fraction of the planet's mass due to intense heating and associated vaporization during an FU Ori outburst, inherently leads to a reduction in the physical size of the planet as well as an increase in its bulk density (as vaporized material preferentially comes from the planetary mantle, leaving an iron core intact).  
We find that depending on the exact system parameters, higher-mass planets in ultra-short period orbits can thus lose a sizeable fraction of their mass, preferentially from the rocky planetary mantle, reducing the physical size and bulk densities of the planets compared to their values at formation.

As shown in \citet{Becker2021}, less massive planets will migrate more readily during FU Ori events. This feedback process - planets lose mass during FU Ori events, which makes them migrate more readily - will generally increase the inwards flux of planets, some of which may end their lives accreted onto their host stars while others end in USP orbits. \editone{As shown in this work, many planets in this range may end their lives through disintegration by thermal effects during the FU Ori event. }

We have presented (Figure \ref{fig:fig2}) a mapping of how the local disk temperature will affect the amount of planet mass retained for a set of planet densities in a single disk environment.  
However, the details of that calculation in the specific outcomes for individual planetary systems cannot be generalized due to natural variations within systems. 
First, the specific mantle composition will alter the efficiency of planetary mass loss, and the sublimation temperature may evolve as material is stripped off the surface of the planet. More detailed modeling \citep[ex:][]{Schaefer2009, Wolf2023} for individual planet compositions may refine the relationship between mass loss and specific planet parameters.
Second, uncertainties in the origin of FU Ori events may alter our understanding of the dynamical environments in which USP planets live. 
For example, it has been proposed \citep{Nayakshin2023, Nayakshin2024} that an evaporating giant planet could be the source of the accretion outbursts seen in FU Ori events. While multi-planet systems containing USP planets and nearby gas giants are not unprecedented \citep{Becker2015}, the alterations to the disk profile in such a case are beyond the scope of our simple model. 
FU Ori outbursts may also have implications on the chemistry available to forming planets at more distant radii of the protoplanetary disk \citep{Calahan2024}.
While short-period, low-mass planets like those we consider may not be present in every system exhibiting FU Ori outbursts, it is possible that in some cases, the high bulk densities of ultra-short-period planets could be tied to mass loss they experienced early in their lifetimes, potentially due to the process we outline in this paper.
\editone{As a final note, the energy-limited mass-loss calculations presented in this paper are qualitative and likely represent a lower limit on the amount of mass lost during FU Ori outburst events. It is possible that a significant fraction of Earth-like planets could be completely vaporized during these events.}

We thank Adolfo Carvalho, Max Goldberg, and Thomas Beatty for useful conversations. \editone{We thank the referee for their careful review of the manuscript and useful suggestions.} 
This research has made use of NASA’s Astrophysics Data System.

\software{pandas \citep{ mckinney-proc-scipy-2010}, matplotlib \citep{Hunter:2007}, numpy \citep{oliphant-2006-guide}, Jupyter \citep{Kluyver:2016aa}}

\bibliography{bib}{}

\begin{thebibliography}{}
\expandafter\ifx\csname natexlab\endcsname\relax\def\natexlab#1{#1}\fi
\providecommand{\url}[1]{\href{#1}{#1}}
\providecommand{\dodoi}[1]{doi:~\href{http://doi.org/#1}{\nolinkurl{#1}}}
\providecommand{\doeprint}[1]{\href{http://ascl.net/#1}{\nolinkurl{http://ascl.net/#1}}}
\providecommand{\doarXiv}[1]{\href{https://arxiv.org/abs/#1}{\nolinkurl{https://arxiv.org/abs/#1}}}

\bibitem[{{Adams} {et~al.}(1988){Adams}, {Lada}, \& {Shu}}]{Adams1988}
{Adams}, F.~C., {Lada}, C.~J., \& {Shu}, F.~H. 1988, \apj, 326, 865, \dodoi{10.1086/166144}

\bibitem[{{Bailey} \& {Zhu}(2024)}]{Bailey2023}
{Bailey}, A.~P., \& {Zhu}, Z. 2024, \mnras, 534, 2953, \dodoi{10.1093/mnras/stae2250}

\bibitem[{{Batygin} {et~al.}(2023){Batygin}, {Adams}, \& {Becker}}]{Batygin2023}
{Batygin}, K., {Adams}, F.~C., \& {Becker}, J. 2023, \apjl, 951, L19, \dodoi{10.3847/2041-8213/acdb5d}

\bibitem[{{Becker} {et~al.}(2021){Becker}, {Batygin}, \& {Adams}}]{Becker2021}
{Becker}, J.~C., {Batygin}, K., \& {Adams}, F.~C. 2021, \apj, 919, 76, \dodoi{10.3847/1538-4357/ac111e}

\bibitem[{{Becker} {et~al.}(2015){Becker}, {Vanderburg}, {Adams}, {Rappaport}, \& {Schwengeler}}]{Becker2015}
{Becker}, J.~C., {Vanderburg}, A., {Adams}, F.~C., {Rappaport}, S.~A., \& {Schwengeler}, H.~M. 2015, \apjl, 812, L18, \dodoi{10.1088/2041-8205/812/2/L18}

\bibitem[{{Bell} \& {Lin}(1994)}]{Bell1994}
{Bell}, K.~R., \& {Lin}, D.~N.~C. 1994, \apj, 427, 987, \dodoi{10.1086/174206}

\bibitem[{{Burrows} {et~al.}(2001){Burrows}, {Hubbard}, {Lunine}, \& {Liebert}}]{Burrows2001}
{Burrows}, A., {Hubbard}, W.~B., {Lunine}, J.~I., \& {Liebert}, J. 2001, Reviews of Modern Physics, 73, 719, \dodoi{10.1103/RevModPhys.73.719}

\bibitem[{{Calahan} {et~al.}(2024){Calahan}, {Bergin}, {van't Hoff}, {Booth}, {{\"O}berg}, {Zhang}, {Calvet}, \& {Hartmann}}]{Calahan2024}
{Calahan}, J.~K., {Bergin}, E.~A., {van't Hoff}, M., {et~al.} 2024, \apj, 975, 170, \dodoi{10.3847/1538-4357/ad78d1}

\bibitem[{{Carvalho} {et~al.}(2024){Carvalho}, {Hillenbrand}, {France}, \& {Herczeg}}]{Carvalho2024}
{Carvalho}, A.~S., {Hillenbrand}, L.~A., {France}, K., \& {Herczeg}, G.~J. 2024, \apjl, 973, L40, \dodoi{10.3847/2041-8213/ad74eb}

\bibitem[{{Essack} {et~al.}(2023){Essack}, {Shporer}, {Burt}, {Seager}, {Cambioni}, {Lin}, {Collins}, {Mamajek}, {Stassun}, {Ricker}, {Vanderspek}, {Latham}, {Winn}, {Jenkins}, {Butler}, {Charbonneau}, {Collins}, {Crane}, {Gan}, {Hellier}, {Howell}, {Irwin}, {Mann}, {Ramadhan}, {Shectman}, {Teske}, {Yee}, {Mireles}, {Quintana}, {Tenenbaum}, {Torres}, \& {Furlan}}]{Essack2023}
{Essack}, Z., {Shporer}, A., {Burt}, J.~A., {et~al.} 2023, \aj, 165, 47, \dodoi{10.3847/1538-3881/ac9c5b}

\bibitem[{{Fegley} \& {Schaefer}(2012)}]{Fegley2012}
{Fegley}, Bruce, J., \& {Schaefer}, L. 2012, arXiv e-prints, arXiv:1210.0270, \dodoi{10.48550/arXiv.1210.0270}

\bibitem[{{Goyal} \& {Wang}(2025)}]{Goyal2025}
{Goyal}, A.~V., \& {Wang}, S. 2025, \aj, 169, 191, \dodoi{10.3847/1538-3881/adb487}

\bibitem[{{Hartmann} \& {Kenyon}(1996)}]{Hartmann1996}
{Hartmann}, L., \& {Kenyon}, S.~J. 1996, \araa, 34, 207, \dodoi{10.1146/annurev.astro.34.1.207}

\bibitem[{{Herbig}(1966)}]{Herbig1966}
{Herbig}, G.~H. 1966, Vistas in Astronomy, 8, 109, \dodoi{10.1016/0083-6656(66)90025-0}

\bibitem[{{Herbig}(1977)}]{Herbig1977}
---. 1977, \apj, 217, 693, \dodoi{10.1086/155615}

\bibitem[{Hunter(2007)}]{Hunter:2007}
Hunter, J.~D. 2007, Computing In Science \& Engineering, 9, 90, \dodoi{10.1109/MCSE.2007.55}

\bibitem[{{Kimura} {et~al.}(2016){Kimura}, {Kunitomo}, \& {Takahashi}}]{Kimura2016}
{Kimura}, S.~S., {Kunitomo}, M., \& {Takahashi}, S.~Z. 2016, \mnras, 461, 2257, \dodoi{10.1093/mnras/stw1531}

\bibitem[{Kluyver {et~al.}(2016)Kluyver, Ragan-Kelley, P{\'e}rez, Granger, Bussonnier, Frederic, Kelley, Hamrick, Grout, Corlay, Ivanov, Avila, Abdalla, \& Willing}]{Kluyver:2016aa}
Kluyver, T., Ragan-Kelley, B., P{\'e}rez, F., {et~al.} 2016, in Positioning and Power in Academic Publishing: Players, Agents and Agendas, ed. F.~Loizides \& B.~Schmidt, IOS Press, 87 -- 90

\bibitem[{{Labdon} {et~al.}(2021){Labdon}, {Kraus}, {Davies}, {Kreplin}, {Monnier}, {Le Bouquin}, {Anugu}, {ten Brummelaar}, {Setterholm}, {Gardner}, {Ennis}, {Lanthermann}, {Schaefer}, \& {Laws}}]{Labdon2021}
{Labdon}, A., {Kraus}, S., {Davies}, C.~L., {et~al.} 2021, \aap, 646, A102, \dodoi{10.1051/0004-6361/202039370}

\bibitem[{{Lam} {et~al.}(2021){Lam}, {Csizmadia}, {Astudillo-Defru}, {Bonfils}, {Gandolfi}, {Padovan}, {Esposito}, {Hellier}, {Hirano}, {Livingston}, {Murgas}, {Smith}, {Collins}, {Mathur}, {Garcia}, {Howell}, {Santos}, {Dai}, {Ricker}, {Vanderspek}, {Latham}, {Seager}, {Winn}, {Jenkins}, {Albrecht}, {Almenara}, {Artigau}, {Barrag{\'a}n}, {Bouchy}, {Cabrera}, {Charbonneau}, {Chaturvedi}, {Chaushev}, {Christiansen}, {Cochran}, {De Meideiros}, {Delfosse}, {D{\'\i}az}, {Doyon}, {Eigm{\"u}ller}, {Figueira}, {Forveille}, {Fridlund}, {Gaisn{\'e}}, {Goffo}, {Georgieva}, {Grziwa}, {Guenther}, {Hatzes}, {Johnson}, {Kab{\'a}th}, {Knudstrup}, {Korth}, {Lewin}, {Lissauer}, {Lovis}, {Luque}, {Melo}, {Morgan}, {Morris}, {Mayor}, {Narita}, {Osborne}, {Palle}, {Pepe}, {Persson}, {Quinn}, {Rauer}, {Redfield}, {Schlieder}, {S{\'e}gransan}, {Serrano}, {Smith}, {{\v{S}}ubjak}, {Twicken}, {Udry}, {Van Eylen}, \& {Vezie}}]{Lam2021}
{Lam}, K. W.~F., {Csizmadia}, S., {Astudillo-Defru}, N., {et~al.} 2021, Science, 374, 1271, \dodoi{10.1126/science.aay3253}

\bibitem[{{Lee} \& {Chiang}(2017)}]{Lee2017}
{Lee}, E.~J., \& {Chiang}, E. 2017, \apj, 842, 40, \dodoi{10.3847/1538-4357/aa6fb3}

\bibitem[{{Livingston} {et~al.}(2024){Livingston}, {Gandolfi}, {Trani}, {Herath}, {Barrag{\'a}n}, {Hatzes}, {Luque}, {Fukui}, {Nowak}, {Palle}, {Hellier}, {Fridlund}, {de Leon}, {Hirano}, {Narita}, {Albrecht}, {Dai}, {Deeg}, {Van Eylen}, {Korth}, \& {Tamura}}]{Livingston2024}
{Livingston}, J.~H., {Gandolfi}, D., {Trani}, A.~A., {et~al.} 2024, Scientific Reports, 14, 27219, \dodoi{10.1038/s41598-024-76490-y}

\bibitem[{{Marley} \& {Robinson}(2015)}]{Marley2015}
{Marley}, M.~S., \& {Robinson}, T.~D. 2015, \araa, 53, 279, \dodoi{10.1146/annurev-astro-082214-122522}

\bibitem[{McKinney(2010)}]{mckinney-proc-scipy-2010}
McKinney, W. 2010, in Proceedings of the 9th Python in Science Conference, ed. S.~van~der Walt \& J.~Millman, 51 -- 56

\bibitem[{{Nayakshin} {et~al.}(2024{\natexlab{a}}){Nayakshin}, {Cruz S{\'a}enz de Miera}, \& {K{\'o}sp{\'a}l}}]{Nayakshin2024}
{Nayakshin}, S., {Cruz S{\'a}enz de Miera}, F., \& {K{\'o}sp{\'a}l}, {\'A}. 2024{\natexlab{a}}, \mnras, 532, L27, \dodoi{10.1093/mnrasl/slae034}

\bibitem[{{Nayakshin} {et~al.}(2024{\natexlab{b}}){Nayakshin}, {Cruz S{\'a}enz de Miera}, {K{\'o}sp{\'a}l}, {{\'C}alovi{\'c}}, {Eisl{\"o}ffel}, \& {Lin}}]{Nayakshin2024b}
{Nayakshin}, S., {Cruz S{\'a}enz de Miera}, F., {K{\'o}sp{\'a}l}, {\'A}., {et~al.} 2024{\natexlab{b}}, \mnras, 530, 1749, \dodoi{10.1093/mnras/stae877}

\bibitem[{{Nayakshin} \& {Lodato}(2012)}]{Nayakshin2012}
{Nayakshin}, S., \& {Lodato}, G. 2012, \mnras, 426, 70, \dodoi{10.1111/j.1365-2966.2012.21612.x}

\bibitem[{{Nayakshin} {et~al.}(2023){Nayakshin}, {Owen}, \& {Elbakyan}}]{Nayakshin2023}
{Nayakshin}, S., {Owen}, J.~E., \& {Elbakyan}, V. 2023, \mnras, 523, 385, \dodoi{10.1093/mnras/stad1392}

\bibitem[{Oliphant(2006)}]{oliphant-2006-guide}
Oliphant, T.~E. 2006, Guide to NumPy, Provo, UT.
\newblock \url{http://www.tramy.us/}

\bibitem[{{Petrovich} {et~al.}(2019){Petrovich}, {Deibert}, \& {Wu}}]{Petrovich2019}
{Petrovich}, C., {Deibert}, E., \& {Wu}, Y. 2019, \aj, 157, 180, \dodoi{10.3847/1538-3881/ab0e0a}

\bibitem[{{Pollack} {et~al.}(1985){Pollack}, {McKay}, \& {Christofferson}}]{Pollack1985}
{Pollack}, J.~B., {McKay}, C.~P., \& {Christofferson}, B.~M. 1985, \icarus, 64, 471, \dodoi{10.1016/0019-1035(85)90069-7}

\bibitem[{{Popham} {et~al.}(1996){Popham}, {Kenyon}, {Hartmann}, \& {Narayan}}]{Popham1996}
{Popham}, R., {Kenyon}, S., {Hartmann}, L., \& {Narayan}, R. 1996, \apj, 473, 422, \dodoi{10.1086/178155}

\bibitem[{{Price} \& {Rogers}(2020)}]{Price2020}
{Price}, E.~M., \& {Rogers}, L.~A. 2020, \apj, 894, 8, \dodoi{10.3847/1538-4357/ab7c67}

\bibitem[{{Rafikov}(2017)}]{Rafikov2017}
{Rafikov}, R.~R. 2017, \apj, 837, 163, \dodoi{10.3847/1538-4357/aa6249}

\bibitem[{{Rodriguez} \& {Hillenbrand}(2022)}]{Rodriguez2022}
{Rodriguez}, A.~C., \& {Hillenbrand}, L.~A. 2022, \apj, 927, 144, \dodoi{10.3847/1538-4357/ac496b}

\bibitem[{{Sanchis-Ojeda} {et~al.}(2014){Sanchis-Ojeda}, {Rappaport}, {Winn}, {Kotson}, {Levine}, \& {El Mellah}}]{SanchisOjeda2014}
{Sanchis-Ojeda}, R., {Rappaport}, S., {Winn}, J.~N., {et~al.} 2014, \apj, 787, 47, \dodoi{10.1088/0004-637X/787/1/47}

\bibitem[{{Schaefer} \& {Fegley}(2009)}]{Schaefer2009}
{Schaefer}, L., \& {Fegley}, B. 2009, \apjl, 703, L113, \dodoi{10.1088/0004-637X/703/2/L113}

\bibitem[{{Serrano} {et~al.}(2022){Serrano}, {Gandolfi}, {Mustill}, {Barrag{\'a}n}, {Korth}, {Dai}, {Redfield}, {Fridlund}, {Lam}, {D{\'\i}az}, {Grziwa}, {Collins}, {Livingston}, {Cochran}, {Hellier}, {Bellomo}, {Trifonov}, {Rodler}, {Alarcon}, {Jenkins}, {Latham}, {Ricker}, {Seager}, {Vanderspeck}, {Winn}, {Albrecht}, {Collins}, {Csizmadia}, {Daylan}, {Deeg}, {Esposito}, {Fausnaugh}, {Georgieva}, {Goffo}, {Guenther}, {Hatzes}, {Howell}, {Jensen}, {Luque}, {Mann}, {Murgas}, {Osborne}, {Palle}, {Persson}, {Rowden}, {Rudat}, {Smith}, {Twicken}, {Van Eylen}, \& {Ziegler}}]{Serrano2022}
{Serrano}, L.~M., {Gandolfi}, D., {Mustill}, A.~J., {et~al.} 2022, Nature Astronomy, 6, 736, \dodoi{10.1038/s41550-022-01641-y}

\bibitem[{{Shakura} \& {Sunyaev}(1973)}]{Shakura1973}
{Shakura}, N.~I., \& {Sunyaev}, R.~A. 1973, \aap, 24, 337

\bibitem[{{Shu} {et~al.}(2007){Shu}, {Galli}, {Lizano}, {Glassgold}, \& {Diamond}}]{Shu2007}
{Shu}, F.~H., {Galli}, D., {Lizano}, S., {Glassgold}, A.~E., \& {Diamond}, P.~H. 2007, \apj, 665, 535, \dodoi{10.1086/519678}

\bibitem[{{Sing} {et~al.}(2024){Sing}, {Rustamkulov}, {Thorngren}, {Barstow}, {Tremblin}, {Alves de Oliveira}, {Beck}, {Birkmann}, {Challener}, {Crouzet}, {Espinoza}, {Ferruit}, {Giardino}, {Gressier}, {Lee}, {Lewis}, {Maiolino}, {Manjavacas}, {Rauscher}, {Sirianni}, \& {Valenti}}]{Sing2024}
{Sing}, D.~K., {Rustamkulov}, Z., {Thorngren}, D.~P., {et~al.} 2024, \nat, 630, 831, \dodoi{10.1038/s41586-024-07395-z}

\bibitem[{{Sonoi} {et~al.}(2019){Sonoi}, {Ludwig}, {Dupret}, {Montalb{\'a}n}, {Samadi}, {Belkacem}, {Caffau}, \& {Goupil}}]{Sonoi2019}
{Sonoi}, T., {Ludwig}, H.~G., {Dupret}, M.~A., {et~al.} 2019, \aap, 621, A84, \dodoi{10.1051/0004-6361/201833495}

\bibitem[{{Tapia} \& {Lizano}(2017)}]{Tapia2017}
{Tapia}, C., \& {Lizano}, S. 2017, \apj, 849, 136, \dodoi{10.3847/1538-4357/aa8f9b}

\bibitem[{{Vazan} {et~al.}(2013){Vazan}, {Kovetz}, {Podolak}, \& {Helled}}]{Vazan2013}
{Vazan}, A., {Kovetz}, A., {Podolak}, M., \& {Helled}, R. 2013, \mnras, 434, 3283, \dodoi{10.1093/mnras/stt1248}

\bibitem[{{Winn} {et~al.}(2018){Winn}, {Sanchis-Ojeda}, \& {Rappaport}}]{Winn2018}
{Winn}, J.~N., {Sanchis-Ojeda}, R., \& {Rappaport}, S. 2018, \nar, 83, 37, \dodoi{10.1016/j.newar.2019.03.006}

\bibitem[{{Wolf} {et~al.}(2023){Wolf}, {J{\"a}ggi}, {Sossi}, \& {Bower}}]{Wolf2023}
{Wolf}, A.~S., {J{\"a}ggi}, N., {Sossi}, P.~A., \& {Bower}, D.~J. 2023, \apj, 947, 64, \dodoi{10.3847/1538-4357/acbcc7}

\bibitem[{{Zhu} {et~al.}(2007){Zhu}, {Hartmann}, {Calvet}, {Hernandez}, {Muzerolle}, \& {Tannirkulam}}]{Zhu2007}
{Zhu}, Z., {Hartmann}, L., {Calvet}, N., {et~al.} 2007, \apj, 669, 483, \dodoi{10.1086/521345}

\bibitem[{{Zhu} {et~al.}(2020){Zhu}, {Jiang}, \& {Stone}}]{Zhu2020}
{Zhu}, Z., {Jiang}, Y.-F., \& {Stone}, J.~M. 2020, \mnras, 495, 3494, \dodoi{10.1093/mnras/staa952}

\bibitem[{Zhu \& Stone(2018)}]{Zhu2018}
Zhu, Z., \& Stone, J.~M. 2018, The Astrophysical Journal, 857, 34, \dodoi{10.3847/1538-4357/aaafc9}

\end{thebibliography}
\bibliographystyle{aasjournal}
\end{document}